\documentclass[10pt,a4paper,twocolumn]{article}
\usepackage{fukugo}
\usepackage{amsmath}
\usepackage{amsfonts}
\usepackage{amssymb}
\begin{document}
%%%%%%%%%%%%%%%%%%%%%%%%%%%%%%%%%%%%%%%
%  Title of your article
%
\title{Generation mixing phenomena for leptons in $e^+e^-$ collisions}
%
%%%%%%%%%%%%%%%%%%%%%%%%%%%%%%%%%%%%%%%
%
%  Author name and address
%

\author{{\large{Noriyuki Oshimo}} \medskip \\ 
  \it{ Department of Physics and Chemistry,} \\ 
  \it{ Graduate School of Humanities and Sciences,       } \\ 
  \it{ Ochanomizu University                             }
       }
%%%%%%%%%%%%%%%%%%%%%%%%%%%%%%%%%%%%%%%
%
%
%
\maketitle
\pagestyle{empty}
\thispagestyle{empty}
\setlength{\baselineskip}{13pt}
%
% Do not change the above four lines
%

%
\def\PL{\left(\frac{1-\gamma_5}{2}\right)}
\def\PR{\left(\frac{1+\gamma_5}{2}\right)}
\def\r2{\sqrt 2}
\def\tanb{\tan\beta}
\def\Usn{\tilde U_\nu}
\def\Un{U_\nu}
\def\la{l_\alpha}
\def\lb{l_\beta}
\def\na{\nu_\alpha}
\def\nb{\nu_\beta}
\def\w{\omega}
\def\sn{\tilde \nu }
\def\m#1{{\tilde m}_#1}
\def\mla{m_{l_\alpha}}
\def\ml#1{m_{l_#1}}
\def\mn#1{m_{\nu_#1}}
\def\Msn#1{M_{\tilde\nu_#1}}

     Mixing of quarks or leptons belonging to different generations
could give a clue to underlying theories for the generations.
We should get phenomenological information about the mixing
as much as possible.
Until recently most experimental data for the mixing were on the
quarks.  
At present, data on the leptons are accumulating from the neutrino
oscillations.  
However, experiments from different aspects are indispensable for 
having a deep insight into the lepton generation mixing.  

\smallskip

     I report the work [NO] which studies measurability of generation 
mixing for the leptons in $e^+e^-$ collisions within the framework of 
the supersymmetric standard model (SSM).   
The SSM is considered a plausible candidate for physics beyond 
the standard model.
This model includes superpartners of quarks and leptons,
which could also be mixed among different generations.
The interaction of charged leptons, sneutrinos, and charginos  
can create two different charged leptons with two
charginos in $e^+e^-$ annihilation.  
The same interaction, on the other hand, induces radiative charged-lepton
decays, which are constrained by non-observation in experiments
to date.
Under these constraints, concentrating on the production of $e$ and $\mu$
with two lighter charginos, we discuss
the possibility of measuring the interaction at $e^+e^-$ colliders.

\smallskip

     The mass eigenstates of the leptons or the sneutrinos are 
generally not the same as their interaction eigenstates.  
Assuming that no superfield for a right-handed neutrino exists 
at the electroweak energy scale, the sneutrinos have a $3\times 3$ 
hermitian mass-squared matrix $\tilde M_\nu^2$.   
The generation mixing of the sneutrinos is traced back to soft 
supersymmetry-breaking terms.  
The mass matrix $M_l$ for the charged leptons is 
proportional to the coefficient matrix of the Higgs couplings.   
For the left-handed neutrinos, we assume a Majorana mass matrix $M_\nu$.   
These mass matrices for the leptons are $3\times 3$.  
The mass eigenstates are obtained by diagonalizing these matrices as 
\begin{eqnarray}
\Usn^\dagger \tilde M_\nu^2\Usn &=& {\rm diag}(\Msn1^2, \Msn2^2, \Msn3^2),  \\ 
U_{lR}^\dagger M_lU_{lL} &=& {\rm diag}(\ml1, \ml2, \ml3),  \\
\Un^T M_\nu \Un &=& {\rm diag}(\mn1, \mn2, \mn3), 
\end{eqnarray}
where $\Usn$, $U_{lR}$, $U_{lL}$, and $\Un$ are unitary matrices.   
The masses of the charged leptons $e$, $\mu$, and $\tau$ are denoted by  
$\ml1$, $\ml2$, and $\ml3$, respectively. 

\smallskip

The interaction Lagrangian for charged leptons $l_\alpha$, 
sneutrinos $\sn_a$, and charginos $\w_i$ is given by
\begin{eqnarray}
\cal L &=& i\frac{g}{\r2}(V_C)_{a\alpha}\sn_a^\dagger\overline{\w_i}
\left[\r2 C_{R1i}^*\PL \right.  \nonumber \\
  &+& \left. C_{L2i}^*\frac{m_{l_\alpha}}{\cos\beta M_W}\PR\right]
l_\alpha \nonumber \\
 &+& {\rm H.c.},
\end{eqnarray}
with $V_C=\Usn^\dagger U_{lL}$.
Here, $C_L$ and $C_R$ denote the unitary matrices for diagonalizing  
the chargino mass matrix which is determined by the SU(2) gaugino 
mass $\m2$, the Higgsino mass $m_H$, and the ratio $\tanb$ of the 
vacuum expectation values for the Higgs bosons.    
The indices $\alpha$ and $a$ for the charged lepton and 
sneutrino, respectively, represent the generations.
The generation mixing is described by $V_C$, which is a $3\times 3$
unitary matrix.
For the independent physical parameters of $V_C$ three mixing angles
$\theta_{12}$, $\theta_{23}$, and $\theta_{13}$ 
and one complex phase $\delta$ can be taken, the other complex phases being
left out by redefinition of particle fields.
We adopt the parametrization in the standard form for the
Cabibbo-Kobayashi-Maskawa matrix.  

\smallskip

     A pair of sneutrinos are created in $e^+e^-$ annihilation 
by the $Z$-boson and chargino exchange diagrams:  
$e^+e^-\to\sn_a\sn_b^*$.  
Assuming that the sneutrinos are heavier than some of the charginos,
the sneutrino can decay into a charged lepton and a chargino
$\sn_a^{(*)}\to\la^{-(+)}\w_i^{+(-)}$.
The generation mixing could be measured by tagging the charged leptons.
However, it is suggested by both theoretical considerations and
experimental constraints derived from radiative charged-lepton decays
that some sneutrino masses are highly degenerated.
Then, the generation mixing has to be analyzed without specifying
the sneutrino generations.
Our analyses are made under an assumption that the sneutrinos
of three generations are produced at the same collision energy and
all the sneutrinos can yield a charged lepton of any generation.

\smallskip

     We discuss the process
$e^+e^-\to\sum_{a,b}\sn_a\sn_b^*\to e^-\mu^+\w^+_1\w^-_1$.
The intermediate sneutrinos are on mass-shell, belonging to
any generations.
This process shows distinctive final states.
The primary leptons $e$ and $\mu$ are produced by two-body decays.
Their energies are large, unless the mass of the lighter chargino is
close to the sneutrino masses.
In addition, they have approximately the same value and are monochromatic
in each rest frame of the decaying sneutrino.
The chargino yields, by three-body decays, two quarks or two leptons
with a neutralino.
Therefore, the final states involve, in each hemisphere, one charged lepton
with a large energy of flat distribution and two jets or one charged lepton,
together with large missing energy-momentum.
The detection of the process will not be difficult.

\smallskip

    The cross section depends on mass differences
of the sneutrinos, as the radiative charged-lepton decays do so.
As the mass difference between $\sn_1$ and $\sn_2$ becomes large,
the cross section increases.
The decay width of $\mu\to e\gamma$ also increases and
becomes too large, thus discarding some ranges of the mass difference.
The mass of $\sn_3$ does not sensitively affect the cross section nor
the decay width.
The constraints from the decays $\tau\to e\gamma$ and $\tau\to\mu\gamma$
are not very stringent, so that large mass differences between
$\sn_1$ and $\sn_3$ and between $\sn_2$ and $\sn_3$ are allowed.
The parameters $\m2$ and $m_H$
do not affect much the mass difference between $\sn_1$ and $\sn_2$ 
allowed by the radiative charged-lepton decays.
However, the cross section varies manifestly
with these parameters, depending on the relative magnitudes of $\m2$
and $m_H$.
For a smaller magnitude of $\m2/m_H$, the SU(2)-gaugino component of
the lighter chargino is larger.
Then, the cross section of $e^+e^-\to\sum_{a,b}\sn_a\sn_b^*$
and thus that of $e^+e^-\to e^-\mu^+\w^+_1\w^-_1$ increase.

\smallskip

   The dependencies of the cross section on the mixing angles of
$V_C$ and $\tanb$ are also manifest.
As the mixing angle increases, the allowed ranges for the mass difference
between $\sn_1$ and $\sn_2$ become narrow,
while the cross section becomes large.
These angle dependencies are primarily determined by $\theta_{12}$.
Different values for $\theta_{23}$ and $\theta_{13}$ do not alter much
the cross section and the allowed mass difference, as long as these
mixing angles are small.
For a larger value of $\tanb$, the allowed region becomes small,
though the cross section does not vary much with it.

\smallskip

    Under the constraints from the presently available experiments,
the cross section of $e^+e^-\to e^-\mu^+\w^+_1\w^-_1$ is larger
than 0.05 fb in sizable regions of the parameter space.
It should be noted that there also exists a charge conjugate process
which has the same cross section.
For the integrated luminosity of 100 fb$^{-1}$, more than ten events
are expected there.
Although possible backgrounds have to be taken into account for
estimating realistically available events, a number of order of
ten would not be insufficient for detection in near future experiments.

\smallskip

In summary, the examination of the process 
$e^+e^-\to e^-\mu^+\w^+_1\w^-_1$ will enable us
to explore the generation mixing peculiar to the SSM.
Moreover, the SSM parameters for the chargino and neutralino sector 
affect non-trivially its cross section.
The study of the generation mixing will give
information on these SSM parameters.

\smallskip

     This work is supported in part by the Grant-in-Aid for
Scientific Research on Priority Areas (No. 16028204) from the
Ministry of Education, Science and Culture, Japan.

\smallskip

\end{document}